\documentclass[useAMS,usenatbib,aas_macros]{mn2e}

\usepackage{lscape}
\usepackage{graphicx}
\usepackage{amssymb}
\usepackage{amsmath}
\usepackage{natbib}
\usepackage{times}
\usepackage{subfigure}
\usepackage{multirow}
\newcommand{\aap}{A\&A}
\newcommand{\mnras}{MNRAS}
\newcommand{\apj}{ApJ}
\newcommand{\apjl}{ApJL}
\newcommand{\apjs}{ApJS}

\catcode`\@=11
\def\gsim{\ifmmode{\mathrel{\mathpalette\@versim>}}
    \else{$\mathrel{\mathpalette\@versim>}$}\fi}
\def\lsim{\ifmmode{\mathrel{\mathpalette\@versim<}}
    \else{$\mathrel{\mathpalette\@versim<}$}\fi}
\def\@versim#1#2{\lower 2.9truept \vbox{\baselineskip 0pt \lineskip 
    0.5truept \ialign{$\m@th#1\hfil##\hfil$\crcr#2\crcr\sim\crcr}}}
\catcode`\@=12
\def\msun{\hbox{$M_\odot$}}
\def\y1{\hbox{${\rm yr}^{-1}$}}
\def\msun{\hbox{$M_\odot$}}

\def\yr-1{\hbox{${\rm yr}^{-1}$}}

\def\t9{\hbox{$t_9$}}

\def\m*{\hbox{$M_{\rm stars}$}}

\def\ho{\hbox{$H_\circ$}}
\def\h50{\hbox{$\ho /50$}}

\title[]{The effect of disk inclination on the Main Sequence of star forming galaxies.}

\author[L. Morselli]{L. Morselli$^{1}$\thanks{E-mail:laura.morselli@tum.de},
A. Renzini $^{2,3}$,
P. Popesso$^{1},$
G. Erfanianfar$^{1}$
\\
$^{1}$Excellence Cluster Universe, Boltzmannstr. 2, Garching bei M\"unchen, 85748, Germany\\
$^{2}$INAF-Osservatorio Astronomico di Padova, Vicolo dell'Osservatorio 5, Padova, I-35122, Italy\\
$^{3}$National Astronomical Observatory of Japan, Mitaka, Tokyo 181-8588, Japan}
\begin{document}

\pagerange{\pageref{firstpage}--\pageref{lastpage}} \pubyear{2016}

\maketitle

\label{firstpage}

\begin{abstract}
We use the Sloan Digital Sky Survey (SDSS) database to explore the
effect of the disk inclination angle on the derived star formation rate (SFR),
hence on the slope and width of the Main Sequence (MS) relation for
star-forming galaxies. We find that SFRs for nearly edge-on disks
are underestimated by factors ranging from $\sim 0.2$ dex for low mass
galaxies up to $\sim 0.4$ dex for high mass galaxies. This results in
a substantially flatter MS relation for high-inclination disks
compared to that for less inclined ones,
though the global effect over the whole sample of star-forming
galaxies is relatively minor, given the small fraction of
high-inclination disks. However, we also find that galaxies with high-inclination
disks represent a non negligible fraction of galaxies populating the
so-called {\it green valley}, with derived  SFRs intermediate  between
the MS and those of quenched, passively evolving galaxies.

\end{abstract}


\section{Introduction}
\label{intro}
The stellar mass and star formation rate (SFR)  of star-forming (SF) galaxies  are tightly correlated with
each other and since  \cite{Noeske07} such a correlation is designed as 
the {\it Main Sequence} (MS) of star-forming galaxies. In a
series of seminal papers \citep{Noeske07,Daddi07,Elbaz07} it was shown  that such tight correlation
persists to at least redshift $\sim 2.5$ with nearly constant slope and
dispersion. Subsequent studies have
confirmed the persistence of a MS relation and extended it all the way to
at least $z\sim 4$ and possibly as much as $z\sim 6$ \citep{Pannella09, Peng10, Rodighiero10,
  Rodighiero11, Rodighiero14, Karim11, Popesso11, Popesso12, Wuyts11, Whitaker12, Whitaker14, Sargent12, 
  Kashino13, Bernhard14, Magnelli14, Speagle14,Steinhardt14}. However,  slope, shape, dispersion and redshift
evolution of the MS relation can vary substantially from
one study to another,  with the logarithmic slope of the relation
ranging  from $\sim 0.4$ up to $\sim 1$, as illustrated by  the compilation in \cite{Speagle14}. Much of these differences can be traced back to the criterion used to select galaxies. For example,  if galaxies
are selected in a SFR-limited fashion, such as in UV- or Far-IR-selected samples, then no MS is recognisable and the SFR
remains roughly constant  with stellar mass \citep{Erb06,Reddy06, Lee13},   because only galaxies with SFR above threshold are recovered, especially at low stellar masses  \citep{ Rodighiero14}.

Besides on the first selection of galaxies, the slope of the MS also depends on how  galaxies are selected as
star forming, e.g., for being bluer than some threshold colour, or for having a specific SFR above a given value, etc. 
To obviate to some of these limitations, \cite{Renzini15} have proposed to use the {\it ridge line} of the 3D distribution of galaxies in the space having for axes SFR, stellar mass and number of galaxies in SFR-mass bins, or, equivalently, taking the mode of the SFR in each mass bin. With this definition, the resulting MS does not depend on a pre-selection of SF galaxies. Using the database of the Sloan Digital Sky Survey (SDSS) data release 7 (DR7)  \citep{Abazajian07}, Renzini \& Peng have derived for the MS of local galaxies a constant slope of $\sim 0.76$, whereas the dispersion may increase somewhat with mass. Using the same database, \cite{Abramson14} have decomposed galaxies in their bulge and disk components, finding that the MS slope gets close to 1 if the SFR is plotted as a function of the {\it disk} mass, as opposed to the total stellar mass. This comes from bulges harbouring very little star formation, if at all, hence contribute mass but not much SFR. Moreover, the fraction of galaxy mass in the bulge component increases steadily with total stellar mass, which then results in a steepening of the MS when the bulge mass is not included. A similar effect (MS slope $\sim 1$) is found for pure disk galaxies at $z\sim1$ \citep{Salmi12}.

Deviations of the MS  from linearity have also been reported, with either a steepening at low masses (e.g., \citealt{Whitaker14}) or flattening at high masses (e.g., \citealt{Whitaker12,Lee15}). These features may arise from a number of effects, such as the inclusion among SF galaxies of objects in which star formation is being  quenched \citep{Renzini15}, or to the development of a passive bulge within the most massive galaxies (e.g., \citealt{Mancini15,Tacchella15, Erfanianfar16}). 

In this paper we explore the effect of the disk inclination on the MS slope and dispersion. Indeed, reddening must be substantially higher for edge-on galaxies compared to face-on ones, and therefore more uncertain the extinction corrections to apply to the derived SFRs. We also study the contamination of inclined disks in the lower envelope of the MS, hence for the population of galaxies that are most likely characterised by green colours, and are found in the valley between the blue cloud and the red sequence in the colour-magnitude diagram (Schawinski et al. 2014, Lee et al. 2012, Leitner et al. 2012, Smethurst et al. 2015).  Here we illustrate the main results of the experiment, having unveiled an inclination effect much stronger than one could have anticipated from the mere extinction correction.

\section{Dataset}
\label{data}
\subsection{SFR and $M_{\star}$}

Also in this paper we use the SDSS-DR7 database, specifically taking SFRs and stellar masses from the MPA-JHU catalogues (http://wwwmpa.mpa-garching.mpg.de/SDSS/DR7/). SFRs have been taken from \cite{Brinchmann04}, where they are derived from the H$\alpha$ luminosity, corrected for reddening. For galaxies with non-detectable H$\alpha$, AGNs or composite galaxies, an estimate of the SFR was derived from the strength of the 4000 \AA\ break (D4000).   

The diameter of the fibers of the SDSS spectrograph is $3''$, hence fibers sample only a fraction of each galaxy, especially at high masses and low redshift. Aperture corrections were then applied to produce the total SFRs listed in the MPA-JHU catalogues. As it will become clear in the following section, aperture corrections play an important role in determining the results of the present investigation and therefore we expand briefly on how they were derived for the MPA-JHU catalogue. The procedure basically follows \cite{Brinchmann04} with one modification. The correction is based on the likelihood of the specific star formation rate for a given set of colours, $P({\rm SFR}/L_{\rm i}/{\rm colour})$, constructing it on a grid of bins of 0.05 mag in $^{0.1}(g-r)$ and of 0.025 mag in $^{0.1}(r-i)$, where SFR/$L_{\rm i}$ is the value of the SFR to $i$-band luminosity ratio inside the fiber, and $^{0.1}(g-r)$ and  $^{0.1}(r-i)$ are the colours, k-corrected to $z=0.1$. The colour of the galaxy outside the fiber was derived by subtracting the fiber luminosities from the total ones, then convolving this estimate of the colour with $P({\rm SFR}/L_{\rm i}/{\rm colour})$ to get a value for SFR/$L_{\rm i}$ outside the fiber. 
\cite{Brinchmann04} warned that the results depend on their main assumption that the distribution of SFR/$L_{\rm i}$ for given $^{0.1}(g-r)$ and $^{0.1}(r-i)$ colours is similar inside and outside the fiber. This assumption may fail for galaxies having a strong bulge component. The mentioned modification consisted in adjusting slightly the SFR  having noted  that the original procedure overestimates the SFR of galaxies with low levels of star formation \citep{Salim07}.
Stellar masses, $M_{\star}$, are taken from  from the SED fits of \cite{Salim07}. 
In the following we make use of the MEDIAN values of SFR and $M_{\star}$, but we have checked that the results are still valid when considering the average (AVG) values.

\subsection{The bulge-disk decomposition: B/T ratio and disk inclination angle}

\begin{figure}
 \centering
 \includegraphics[width=0.45\textwidth, keepaspectratio]{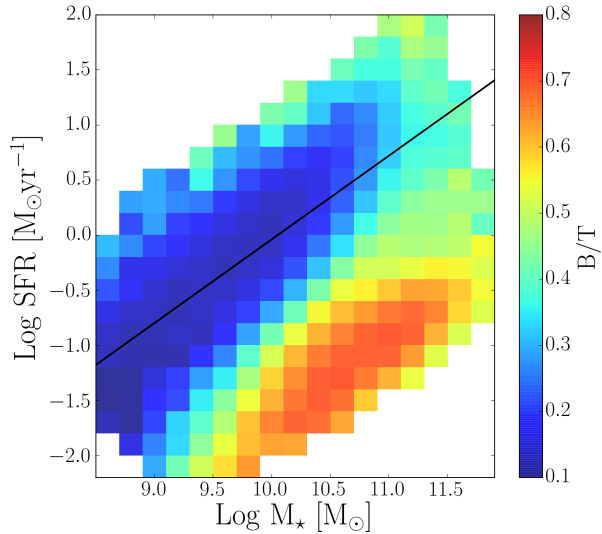}
\caption{ The SFR vs. stellar mass plane of galaxies with $0.02<z<0.2$, color-coded by the average B/T value in each bin. The bins are 0.2 dex wide in Log(SFR) and 0.2 dex wide in Log($M_{\star}$), and each bin host a minimum number of 20 galaxies. The B/T values are taken from the Simard et al. (2011) catalogue, and are computed from $r-$band images. The black solid line is the MS relation computed by Renzini $\&$ Peng (2015). The B/T has its minimum on the MS, and it increases in the upper envelope of the MS, and for green valley and passive galaxies. Restricting the analysis to B/T$\le0.5$ allows us to avoid from the sample passive galaxies and focus on MS and green valley galaxies.}
\label{f0}
 \end{figure}
 
 \begin{figure*}
 \centering
 \includegraphics[width=0.8\textwidth, keepaspectratio]{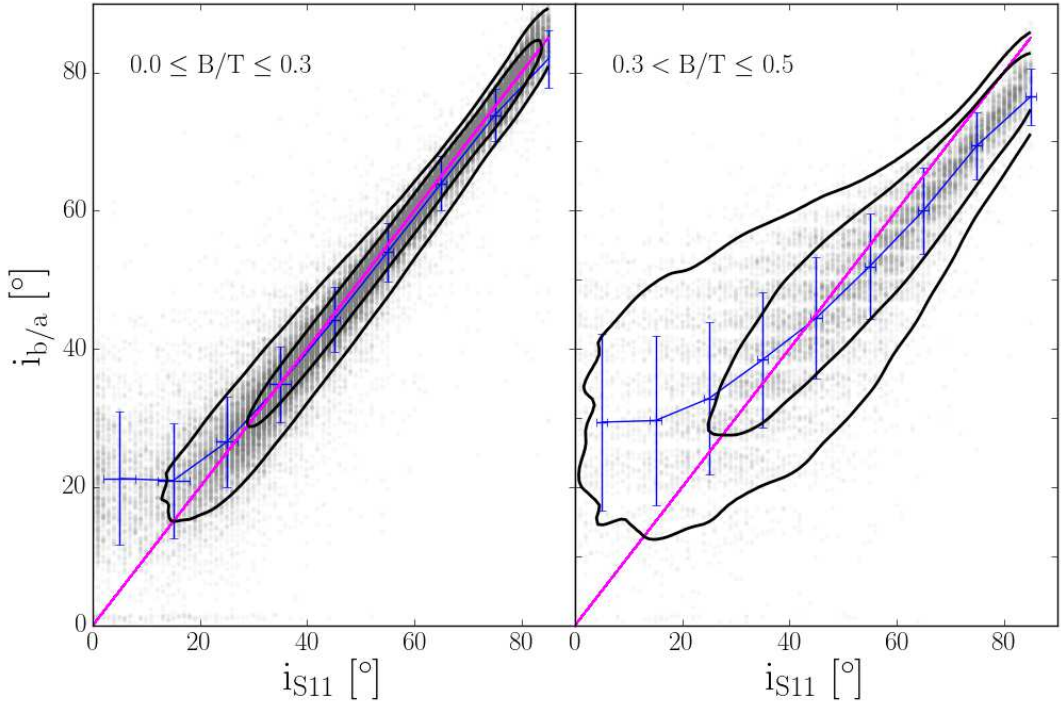}
\caption{Disk inclination angle in the Simard et al. (2011) catalogue ($i_{S11}$) versus the inclination angle derived from the minor-to-major axis ratio of SDSS, computed from a single $n=1$ S\'ersic profile ($i_{b/a}$). In the left panel, the subsample of galaxies with B/T$\le0.3$ is shown, while the right panel refers to galaxies with 0.3$<$B/T$\le$0.5. The magenta line is the  1-to-1 relation, while the black contours encircle the 68$\%$ and $95\%$ of the galaxy population. The blue points mark the median $i_{b/a}$ value in intervals of 10$^{\circ}$ in $i_{S11}$. The width of the errorbars in the $y$-direction is the standard deviation of the $i_{b/a}$ distribution in each bin while the width in the $x$-direction is the median of the errors in $i_{S11}$ in each 10$^{\circ}$ wide bin. In the lowest B/T bin, the correlation $i_{S11}$ and $i_{b/a}$ is extremely tight and consistent with the 1-to-1 relation. For galaxies with 0.3$<$B/T$\le$0.5, the scatter of the $i_{S11}$ - $i_{b/a}$ relation increases,  but the slope is still consistent with $m=1$.} 
\label{f1}
 \end{figure*}

We use the \citet[][S11 hereafter]{Simard11} bulge-disk decomposition of galaxies in the SDSS DR7 photometric catalogue, done with the GIM2D code (Simard et al. 2002). In particular, we make use of the decomposition done with a bulge+disk model, where the bulge S\'ersic index is fixed to $n=4$ and the disk component has $n=1$. This decomposition gives, together with other parameters, the B/T ratio in the $r$ and $g$ bands, and the disk inclination angle, that is $i =0^\circ$ for face-on galaxies and $i=90^\circ$ for edge-one ones. Other quantities like the B/T ratio within the SDSS fiber in the $g$ and $r$ bands are also given. In addition, in the S11 catalogue the probability that a double component disk+bulge model is not required to describe a galaxy is stored in the $P_{PS}$ parameter. $P_{PS}$ is a useful quantity to select trustable disk+bulge galaxies.

Morselli et al. (2016, submitted to MNRAS; hereafter M16) carried out a comparison of the morphological classification of S11 with others, specifically Galaxy Zoo \citep{Lintott11,Willett13} and BUDDA \citep{DeSouza04}, and found that the three morphological classifications are in good agreement, especially for low B/T values.  

One of the first step of this study is to understand the reliability of disk parameters, specifically the inclination angle, for the galaxies we are interested in. In fact, it is straightforward to understand that disk parameters are more easy to retrieve from disk-dominated galaxies, while for bulge-dominated ones it is not trivial due to the presence of a luminous bulge. The aim of this work is to study the effect of disk inclination on star-forming galaxies and the contamination of inclined disks in the green valley region. Hence, it is useful to try to understand wether the disk inclination angle is reliable for the galaxies we are interested in. M16 shows that for galaxies in the range $0.02<z<0.1$ the MS region of the SFR-$M_{\star}$ plane is dominated by disk galaxies, with B/T values on the MS up to 0.3 for the most massive galaxies. The average B/T values increase when moving from the MS towards its upper envelope and towards the green valley and quiescent region. Here, for completeness, we repeat the same analysis but considering galaxies in the redshift range of this work: $0.02<z<0.2$. Figure \ref{f0} shows the SFR-$M_{\star}$ plane divided in bins that are 0.2 dex wide in Log($M_{\star}$), and 0.2 dex wide in Log(SFR), where each bin is color-coded as a function of the average B/T ratio, as indicated by the color bar on the right. Here and trough the paper, we make use of the B/T ratio computed from the $r$ image. In Figure \ref{f0}, the MS is indicated by the solid black line, and its equation is taken from R$\&$P15. It is clear that star forming galaxies are on average disk-dominated, with an increase of the average B/T value on MS at higher stellar mass. It is evident that galaxies with SFRs up to $\sim$ 1 dex below the MS have B/T values that are, on average, smaller than 0.5. Galaxies with the lowest SFRs have, on average, higher B/Ts, due to the presence of a mixed population of red spirals and red ellipticals in the passive region on the SFR-$M_{\star}$ plane (M16).

 \begin{figure*}
 \centering
 \includegraphics[width=0.48\textwidth, keepaspectratio]{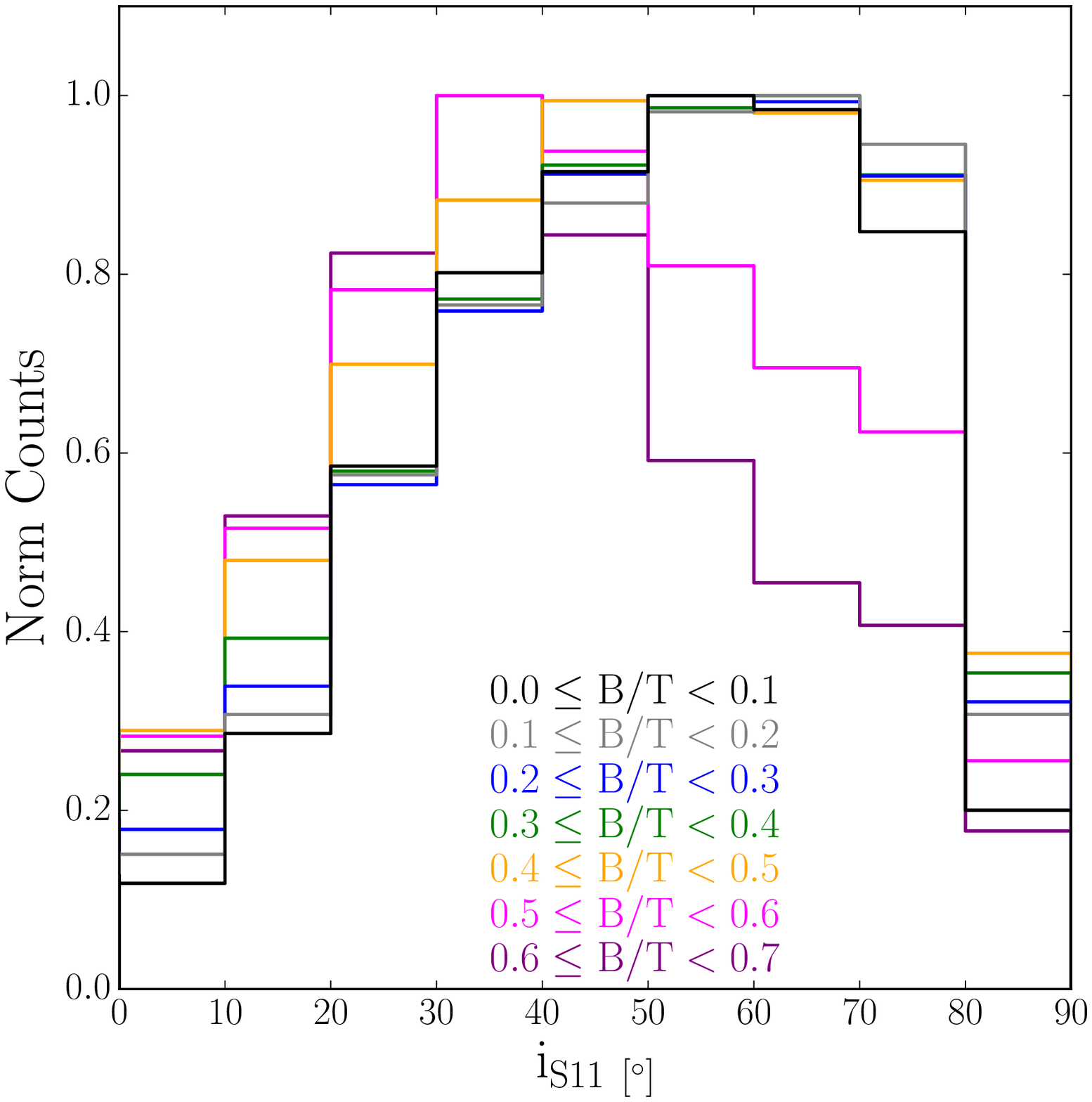}
  \includegraphics[width=0.48\textwidth, keepaspectratio]{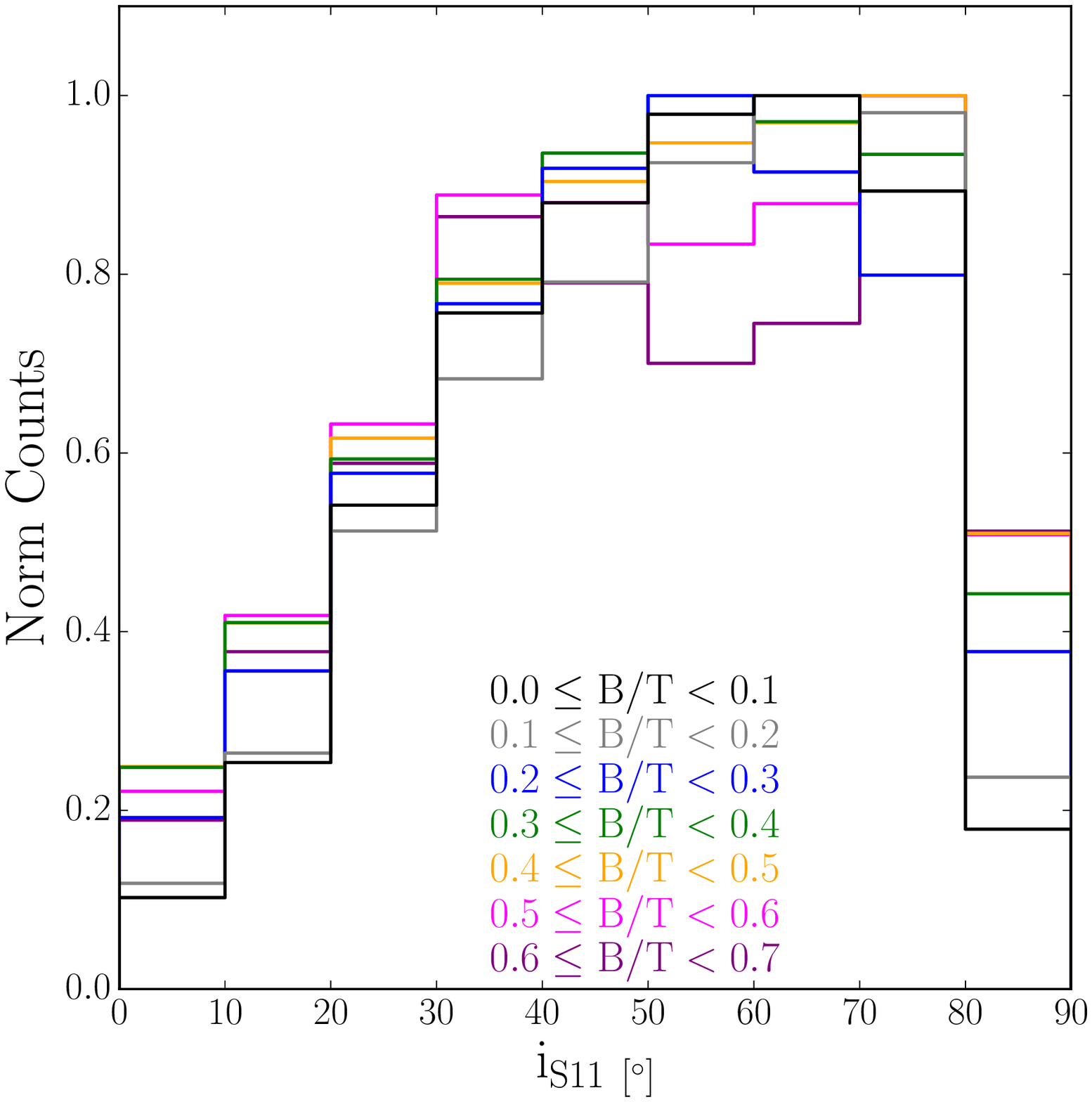}
\caption{Distributions of the disk inclination angle in the S11 catalogue in different bins of B/Ts, indicated by different colours: 
0.0$\le$B/T$<$0.1 in black, 0.1$\le$B/T$<$0.2 in gray, 0.2$\le$B/T$<$0.3 in blue, 0.3$\le$B/T$<$0.4 in green, 0.4$\le$B/T$<$0.5 in orange, 0.5$\le$B/T$<$0.6 in magenta, and 0.6$\le$B/T$<$0.7 in purple. The disk inclination angles for galaxies with B/T$>$0.5 is clearly biased by the presence of the luminous bulge, and hence not reliable.   }
\label{f2}
 \end{figure*}

We continue our analysis with the comparison of the disk inclination angle from the S11 catalogue, $i_{S11}$, and the inclination angle given by the minor-to-major axis ratio ($b/a$) of the single S\'ersic-index n=1 model that fits the surface brightness distribution of a galaxy, as from the SDSS pipelines ($i_{b/a}={\rm arccos}(b/a)$). This is done in two bins of B/T ratio, to access the reliability of the disk inclination angle in S11. In particular, we divide our sample in galaxies with B/T$\le0.3$ (left panel of Figure \ref{f1}), and 0.3$<$B/T$\le$0.5 (right panel), to check trends with morphology. The contours in both panels of Figure \ref{f1} encircle the areas where the 68$\%$ and 95$\%$ of the population are found. The magenta solid line marks the 1-to-1 relation, while the blue points are the median $i_{b/a}$ in intervals of 10$^{\circ}$ in $i_{S11}$. The errorbars in the $y$-direction are given from the standard deviation of the $i_{b/a}$ distribution in each bin, while the median of the errors in $i_{S11}$ in each 10$^{\circ}$ wide bin give the error bar extension in the $x$-direction. As expected, for disk dominated galaxies the correlation between $i_{S11}$ and $i_{b/a}$ is consistent with the 1-to-1 relation and the scatter is very small. This means that the GIM2D code is perfectly able to retrieve reliable disk inclination angles, also when a small bulge is present. For $i_{S11}<20^{\circ}$ there is a flattening to a value of $\sim20^{\circ}$ of the disk inclination angle given from the axis ratio. For galaxies with a larger bulge component, 0.3$<$B/T$\le$0.5, the correlation is still consistent with a 1-to-1 relation, but the scatter is larger and the slope slightly smaller than 1. For galaxies with $i_{S11}>50^{\circ}$, the disk inclination angle $i_{b/a}$ is always smaller than $i_{S11}$, and the difference increases for increasing inclination. It is straightforward to understand that this behaviour is the consequence of the fact that a single $n=1$ S\'ersic model is not anymore a good approximation of the surface brightness profile in this range of B/Ts. Especially in the case of nearly edge-on galaxies, even if the disk axis ratio is small (and hence the disk highly inclined), the presence of a luminous spherical bulge in the centre will result in an increase of the axis ratio that comes out from a single S\'ersic fit. The analysis shown in Figure \ref{f1} tells us that the disk inclination angle of the S11 catalogue is extremely reliable at low B/T values, but we further tested it as a function of morphology. A perfect tool would retrieve the same distribution of disks inclination angles, independently on how dominant the bulge component is. The left panel of figure \ref{f2} shows the distributions of the disk inclination angle $i_{S11}$ for different subsamples of galaxies, defined by their B/T ratio, as indicated by different colours. For B/T$<0.4$ the distributions are similar, with a peak between 50 and 70 degrees. For galaxies with 0.4$\le$B/T$<$0.5, the distribution is broader, but the peak is still between 50 and 70 degrees. The distribution of disk inclination angles drastically changes when considering galaxies with more dominant bulges: the peak moves between 30 and 40 degrees, due to the presence of a massive, luminous bulge.
 This behaviour can be the result of a trend of the GIM2D code, well discussed in S11, of finding spurious disks in bulge-dominated galaxies. Hence, we repeated the same analysis of the distribution of disk inclination angles as a function of the galaxy B/T, just considering galaxies for which the statistical F-test confirms the need of a double component model to fit the galaxy surface brightness profile. This is done, as suggested by S11, by considering only galaxies with $P_{PS}<$0.32, and it is shown in the right panel of Fig. \ref{f2}. We first notice that the peak in the $i_{S11}$ distribution for galaxies with B/T$<$0.1 slightly shifts between 40$^{\circ}$ and 50$^{\circ}$; this is because when we restrict the sample to galaxies with $P_{PS}<$0.32 we exclude all the pure disks that do not host a bulge, that have a peak in $i_{S11}$ between 60$^{\circ}$ and 70$^{\circ}$. For galaxies with 0.1$\le$B/T$<$0.4 and $P_{PS}<$0.32, the distributions of $i_{S11}$ are very similar to the distributions shown in the left panel. When 0.4$\le$B/T$<$0.5, the distribution is broader and the peak is found at slightly larger angles than for smaller B/Ts. The distributions become nearly flat between 30$^{\circ}$ and 70$^{\circ}$, with a peak between 70$^{\circ}$ and 80$^{\circ}$ for galaxies with B/T$\ge$0.5. This could be interpreted as the fact that it is easier for the code to retrieve the presence of the disk in a bulge dominated galaxy when the disk is nearly edge-on.

This tests tell us that the code is able to retrieve meaningful disk inclination angles up to B/T $\simeq$ 0.5 for the whole sample of galaxies. When a galaxy is strongly bulge dominated, the code tends to find spurious disks and to give them low inclination angles, most likely due to the fact that the axis ratio is controlled by the intrinsic shape of the spheroidal bulge.

We conclude that the disk inclination angle in S11 gives us the possibilities to study the disk inclination effects also on those galaxies that have a certain, non dominant (B/T$\leq0.5$,) bulge component, giving us a more realistic estimate of the angle with respect to the disk inclination angle from the axis ratio obtain from a single Sersic model fit. Those galaxies dominate the star forming and green valley regions of the SFR-$M_{\star}$ plane, the region we are interested in this work.

 \begin{figure}
 \centering
 \includegraphics[width=0.48\textwidth, keepaspectratio]{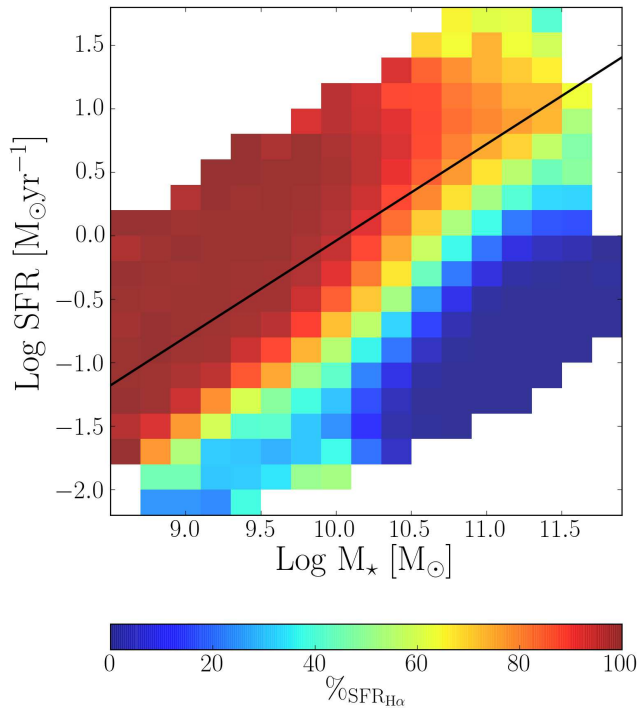}
\caption{The SFR-$M_{\star}$ plane color coded by the fraction of galaxies for which the SFR was derived from the H$\alpha$ flux, as opposed from D4000, as indicated by the color bar at the bottom. The black solid line is the MS computed by R$\&$P15. A fraction always larger of 70$\%$ of MS galaxies has SFR from H$\alpha$. The fraction decreases when moving from the MS towards its lower envelop, but it is always larger than $\sim50\%$ for the galaxies we focus on in this work.}
\label{f3}
 \end{figure}

\subsection{The final catalogue}

 \begin{figure}
 \centering
 \includegraphics[width=0.48\textwidth, keepaspectratio]{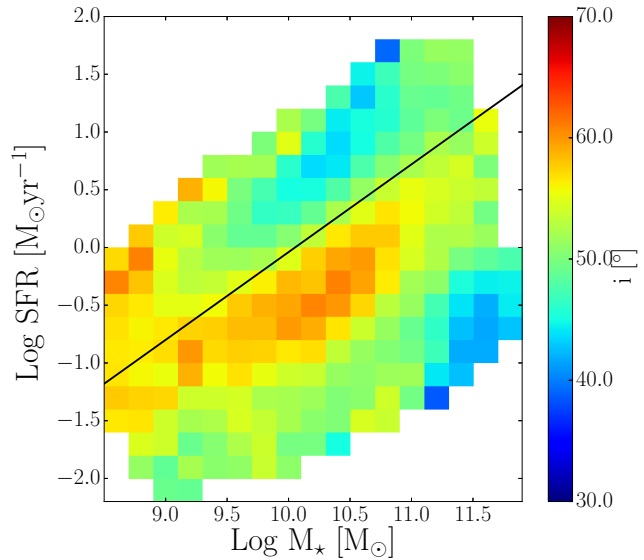}
\caption{The SFR vs. stellar mass plane of galaxies with $0.02<z<0.2$ and B/T$\le0.5$, color-coded by the average disk inclination angle ($i$)  in the bin. The bins are 0.2 dex wide in Log(SFR) and 0.2 dex wide in Log($M_{\star}$), and each bin host a minimum number of 20 galaxies (as in Figure \ref{f0}).  The disk inclination angles are taken from the Simard et al. (2011) catalogue. The black solid line is the MS relation computed by Renzini $\&$ Peng (2015). The disk inclination angle is not randomly distributed in the plane, as expected if there would be no biases in the determination of the SFRs. On average, galaxies in the lower envelope of the MS relation are characterised by more inclined disks.}
\label{f5}
 \end{figure}

In this work we consider only galaxies with Petrosian extinction-corrected magnitude $m_{\rm Petro_r,corr}\le17.77$, which is the completeness limit of the Main Galaxy spectroscopic sample, $0.02<z<0.2$. We excluded active galactic nuclei (AGNs) from the sample, since  SFR rate estimates could be affected by AGN emission lines. To do so, we consider only objects whose spectra are classified as {\it galaxy}, excluding galaxies with {\it SUBCLASS} of AGN; these criteria led to the exclusion of  $\sim4\%$ of sources which otherwise would have fulfilled the other selection criteria. Also, we focus our analysis on galaxies that have $8.5\le {\rm Log(M_{\star}/M_{\odot})}\le11.25$.  The SDSS DR7 galaxies that satisfy all those criteria are $\sim$520,000.  
Given  the tests on the reliability of the disk inclination angle as measured in S11 and of the distribution of the B/T ratio in the SFR-$M_{\star}$ plane, we focus mainly on the $\sim$320,000 galaxies with B/T$\le 0.5$. However, we will  briefly mention also how the derived SFRs correlate with the axis ratio of  bulge-dominated (B/T$>0.5$) galaxies.

To correct for incompleteness at low stellar masses, we applied the $V/V_{max}$ correction, where $V_{max}$ is the maximum volume at which a galaxy with a given H$\alpha$ flux would still be detected with S/N$\ge 3$. We underline here that this correction, based on H$\alpha$, is valid only for galaxies whose SFR comes from the H$\alpha$ flux. This is a good approximation since the SFR comes from H$\alpha$ for the large fraction of low stellar mass galaxies for which incompleteness needs to be taken into account. Figure \ref{f3} shows the fraction of galaxies in our sample with SFR from H$\alpha$, as a function of their position in the SFR-$M_{\star}$ plane. This fraction is always very large on the MS, with average values always above 70$\%$. For galaxies with M$_{\star}<10^{10}M_{\odot}$ this fraction is $\sim100\%$ on the MS and its upper envelope , while it decreases to $50\%$ at $\sim1$ dex below the MS. Hence we can safely say that the incompleteness correction accounts for the majority of galaxies in the region of the SFR-$M_{\star}$ plane in which we are interested. At higher stellar masses, the fraction of star forming galaxies with SFR from H$\alpha$ decreases on the MS, and more rapidly when moving from the MS to its lower envelop, but here the spectroscopic completeness is high, hence the correction is not needed. It is nevertheless important to underline that the SFR from D4000 should be less affected by dust extinction than SFR from H$\alpha$. The systematic change along the MS in the fraction of SFRs derived from H$\alpha$ (or D4000) may introduce a bias in e.g., the derived slope of the MS or in the population of the green valley, an aspect we do not investigate further in the present paper.

\section{Inclination effects on the SFR-$M_{\star}$ relation.}

\begin{figure*}
 \centering
 \includegraphics[width=0.95\textwidth, keepaspectratio]{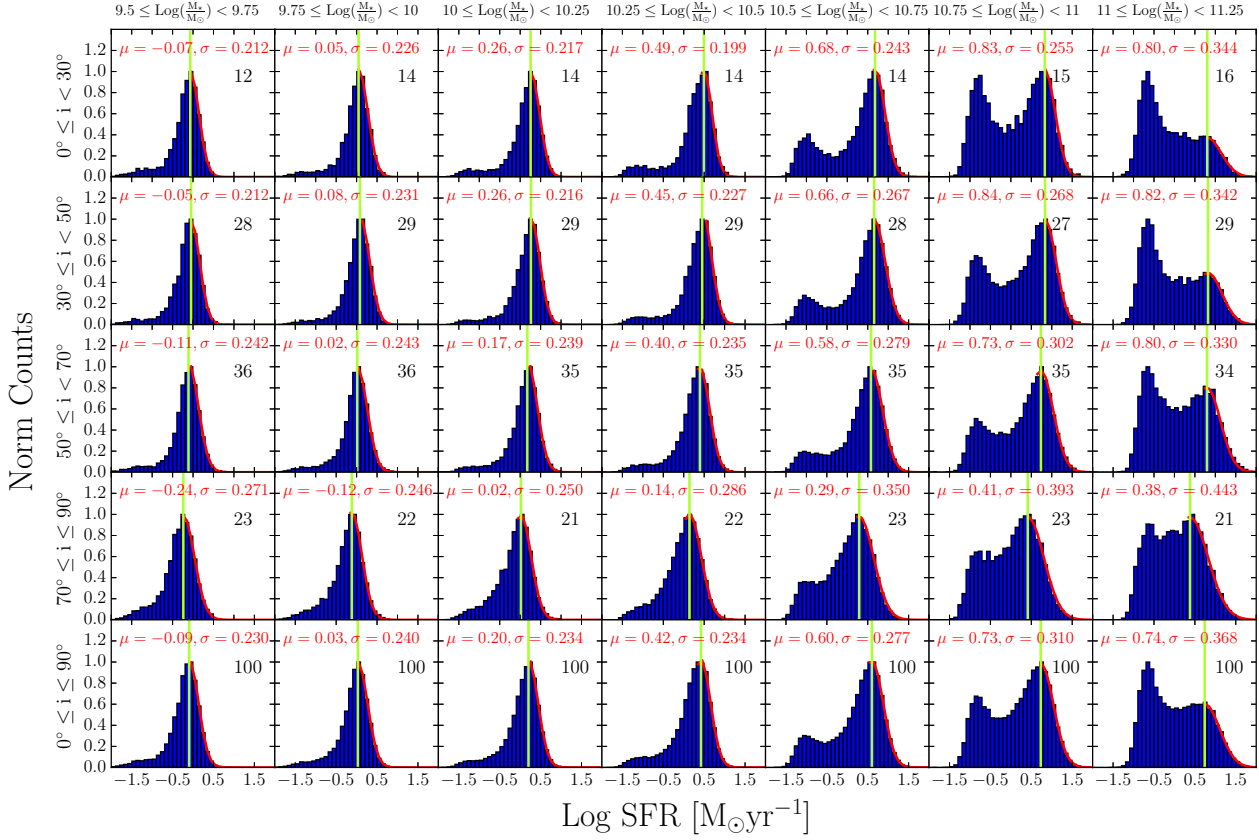}
\caption{Total SFR distributions in bin of stellar mass (different columns, increasing from left to right) and in bins of disk inclination angle (different rows, increasing from top to bottom), for galaxies with B/T$\le$0.5. The bottom row shows the whole galaxy sample, i.e. 0$^{\circ}\le i \le 90^{\circ}$. The right part of the SF distribution has been fitted with a gaussian (in red), and the best fit $mean$ and $dispersion$ values are written in each panel in red. In each panel, also the fraction of galaxies is shown. The green lines mark the position of the mean value of the best-fit gaussian. The number in each panel is the percentage of galaxies in that bin of disk inclination angle and total stellar mass. For a given stellar mass bin, the peak of the SF distribution shifts at lower SFR values with increasing disk inclination angle, and the dispersion of the distribution increases.  }
\label{f6}
 \end{figure*}

If we had a perfect tool to derive the SFR of galaxies the results should be independent on their aspect with respect to the observer.  In practice, however, our tools are not perfect. Edge-on galaxies are obviously more affected by reddening and extinction than face-on ones and reddening corrections are therefore more uncertain. In addition, self-extinction is not the only disk-inclination dependant effect that can result in a biased SFR estimate. In fact, for galaxies that have intermediate morphology, the inclination of the disk translates on the portion of the disk that enters the fiber, as we will discuss later. For example, a face-on galaxy with B/T$\sim$0.5 at low redshift, will have a fiber SFR that is substantially given by the bulge, and hence it is likely to be extremely low. On the contrary, when observing the same galaxy when the disk is edge-on, the fiber SFR will also have the contribution of the disk, even if it might be underestimated due to its internal extinction. To obtain the total SFR of a galaxy, a substantial aperture correction is applied, which relies on simplifying assumptions that may introduce errors and biases.

Here we want to study the effects of disk inclination on the determination of the galaxy SFR and, as a consequence, on the slope, shape and dispersion  of the MS of SF galaxies. We start our analysis by showing how the disk inclination angle varies along the SFR-$M_{\star}$ plane for the galaxies in our sample. Results are shown in Figure \ref{f5}, where each bin in stellar mass and SFR has been colour-coded as a function of the average value of the disk inclination angle, $i_{S11}$, in each bin. The black solid line is the MS of R$\&$P15. It is clear that the highest average disk inclination angles run parallel to the MS, but offset by about 0.5 dex. Galaxies that are above the MS are characterised by lower average disk inclination angles. For SFRs $\sim$1 dex below the MS, the average inclination angle is between $\sim45^{\circ} - 55^{\circ}$, i.e. in the range of the most probable disk inclination.  The behaviour of the disk inclination angle in the SFR-$M_{\star}$ plane clearly reveals the presence of biases in the estimate of the SFRs.

To further investigate the dependence of the SFR estimate on the disk inclination angle, we plot in Figure \ref{f6} the distribution of the SFR in different bins of stellar mass (0.25 dex wide), increasing  from left to right, and in different bins of disk inclination angle (from top to bottom, as indicated), for galaxies with B/T$\le$0.5. The entire sample ($0^\circ\le i\le 90^\circ$) is illustrated in the bottom row. All the panels show the classical bimodal distribution for SF and for quenched galaxies, with the quenched peak becoming progressively more prominent with increasing mass. The star-forming peaks have been fitted with a gaussian that is shown in red, and the resulting mean ($\mu$) and dispersion ($\sigma$) values are reported  in each panel together with the fraction of galaxies in each bin. We fit only the right part of the SF distribution, to avoid a shift of the mean due to the likely presence of quenching galaxies in the {\it green valley}.  We can identify different effects of the inclination. For each stellar mass bin, the peak  of the SFR distribution moves to lower values with increasing inclination angle. This difference in the peak position between lowest and highest disk inclination bins is $\sim 0.2$ dex for galaxies with  ${\rm Log(M_{\star}/M_{\odot})}\le10.25$, and it increases with increasing stellar mass, up to $\sim0.4$ dex for the most massive galaxies. Also, for a fixed stellar mass bin the dispersion of the star-forming gaussian increases with increasing disk inclination angle. This effect is relatively small (up to 0.1 dex in the most massive galaxies). A consequence of the shift to lower values of the SF peak is the progressive partial filling of the green valley, which almost disappears for the most massive, nearly edge-on galaxies. When considering the whole sample, irrespective of inclination, the distributions closely resembles those for the most probable disk inclination, i.e. $\sim 50^\circ$. 

 For the most massive galaxies, ${\rm Log(M_{\star}/M_{\odot})}>10.5$, a change in the relative strength of the two peaks in the SFR distribution, the quenched and star forming ones, is also noticeable. With the  disk inclination angle increasing from 0$^{\circ}$ to 90$^{\circ}$, the peak of passive galaxies becomes less prominent, in favour of a more significant SF peak.  More massive MS galaxies are, on average, more bulge dominated, hence the fiber samples different portions of the disk depending on the disk inclination angle. For nearly face-on galaxies the fiber may sample just the passive bulge and the whole galaxy is assigned a low SFR.  With  increasing disk inclination angle, the portion of the disk that enters the fiber increases, so the galaxy is recognized as star-forming and assigned a high SFR.  This is further discussed in Section 4.1.

 \begin{figure*}
 \centering
 \includegraphics[width=0.95\textwidth, keepaspectratio]{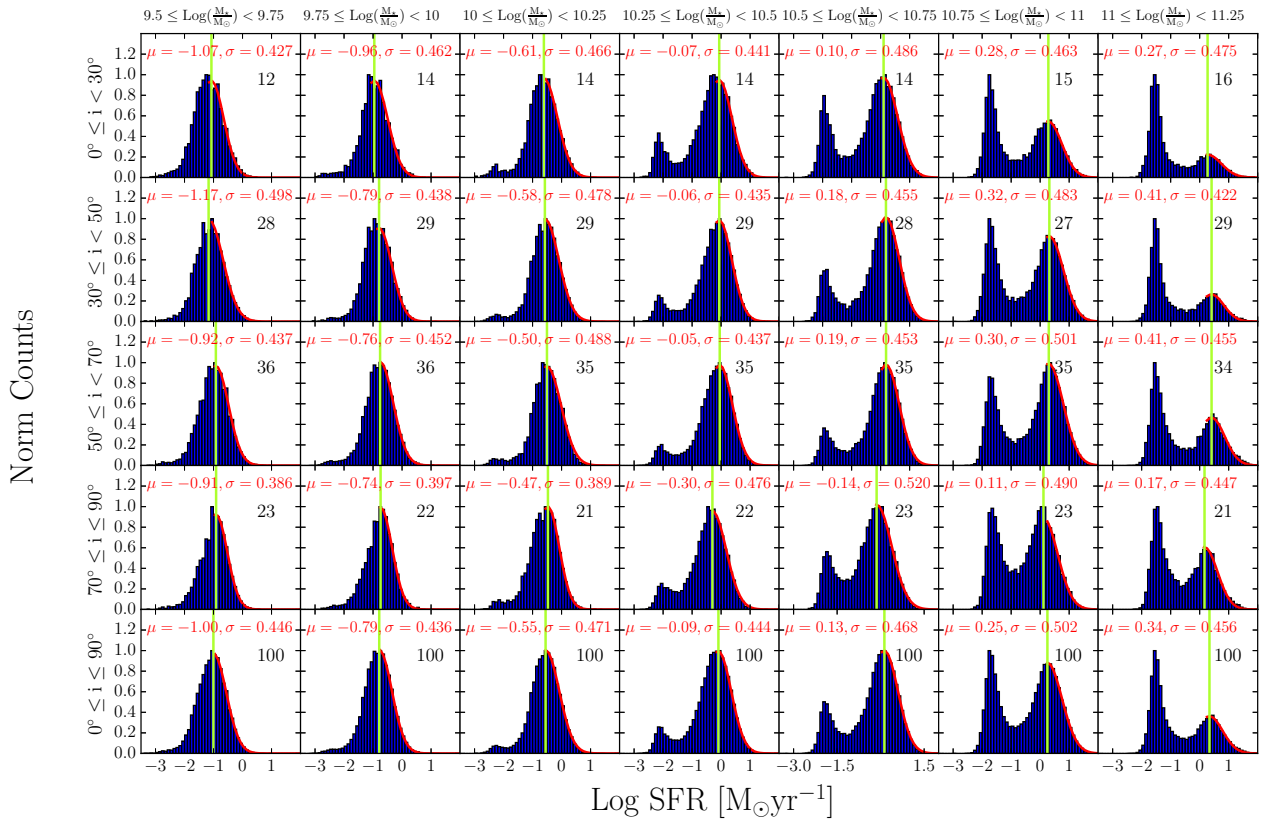}
 \caption{Same as in Fig. \ref{f6}, but considering the fiber SFR, not aperture-corrected.}
\label{f7}
 \end{figure*}

As discussed in Section \ref{data}, a key step in determining the SFR of a galaxy is the aperture correction, hence we now check whether the shift in the SF peak is present also when considering the fiber values, i.e., the SFR measured within the portion of the galaxy sampled by the fiber, without applying the aperture correction. Figure \ref{f7}, analogues to Figure \ref{f6}, shows the results. We can identify two different behaviours of the fiber SFR with disk inclination, depending on the stellar mass. 
\begin{itemize}
\item Galaxies with M$_{\star} < 10^{10.25}M_{\odot}$.
The trend reverses with respect to the one in Figure \ref{f6}: the peak of the SF distribution shifts to slightly higher values for increasing disk inclination angle, whereas it shifts to slightly lower values in Figure \ref{f6}. 

\begin{figure*}
 \centering
 \includegraphics[width=0.95\textwidth, keepaspectratio]{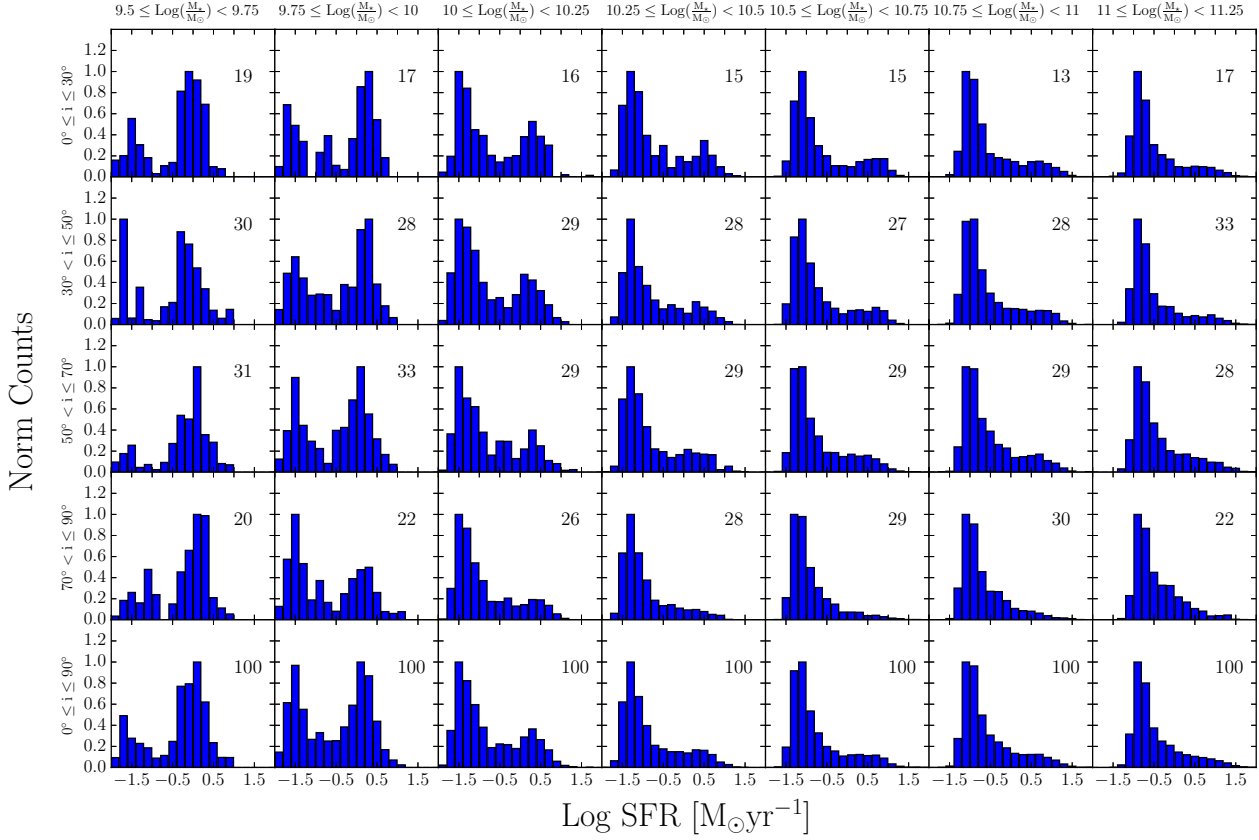}
\caption{Total SFR distributions in bin of stellar mass and disk inclination angle ($i_{S11}$) analogues to Figure \ref{f6}, but for galaxies with 0.5$<$B/T$<$0.9 and $P_{PS} <$0.32 (to exclude spurious disks from the subsample of bulge-dominated galaxies).}
\label{SFR_histogram.pdf}
 \end{figure*}

\item Galaxies with M$_{\star} \ge 10^{10.25}M_{\odot}$.  
We notice that the fiber SF peaks at lower SFR values for nearly face-on disks ($i<30^{\circ}$) than for higher disk inclination angles (30$^{\circ}<i<$70$^{\circ}$). This is because the bulge component of MS galaxies becomes more prominent for increasing stellar masses (see Figure \ref{f0}), and hence the fiber SFR of a galaxy with a face-on disk is affected by the passive bulge component. This is confirmed by the fact that, for the most massive galaxies with disk inclination angle between 0$^{\circ}$ and 30$^{\circ}$, the passive peak is more predominant in the fiber SFR distribution (Figure \ref{f7}) than in total SFR distribution (Figure  \ref{f6}). \footnote{On the contrary, for nearly face-on galaxies with lower stellar masses, the shape of the SFR distribution does not change when considering the total SFR instead of the fiber value, because these galaxies are either pure disks or have very low B/T values. }
For increasing disk inclination angles ($i>70^{\circ}$), the SF peak shifts to smaller values ($\sim$0.2 dex), and is more predominant than for lower disk inclination angles. We infer that since a larger portion of the disk enters the fiber, a larger number of galaxies is identified as star-forming, but the dust correction results in an underestimate of the SFR. In fact, morphology alone can not account for this as dust correction shifts the fiber SF peak to lower values (quantitatively less pronounced), even for nearly edge-on pure-disk massive galaxies.
\end{itemize}
The varying trends in the position of the fiber SFR peak as a function of disk inclination, for galaxies with different stellar masses, can be attributed to their dust content. As shown in Brinchmann et al. (2004),  with increasing stellar mass galaxies have a higher dust content, hence the trend of self absorption with disk inclination is more important for more massive galaxies.

\indent For completeness, we repeated the same analysis for the bulge-dominated galaxies, with 0.5$<$B/T$<$0.9. We know that the S11 catalogue is affected by the presence of spurious disks in bulge-dominated galaxies, and that those disks are given, on average, lower inclination angles. For this reason, we restrict the sample for this qualitatively analysis only to galaxies that have a secure disk-component, i.e. $P_{PS}<0.32$. Figure \ref{SFR_histogram.pdf} shows the total SFR histograms in mass and axis ratio bins, analogues to Figure \ref{f6}, but now for bulge-dominated galaxies. Not unexpectedly, the quenched peak is dominant in virtually all bins, including the low mass ones, contrary to Figure \ref{f6} where passive galaxies are almost completely absent in the low-mass bins. We can observe, especially for more massive galaxies, a progressive filling of the green valley region with increasing disk inclination. This progressive filling is due to the shift at lower values of the peak of the SF gaussian, that at high disk inclinations and stellar masses it is not distinguishable anymore from the tail of the passive gaussian.  

\begin{figure}
 \centering
 \includegraphics[width=0.44\textwidth]{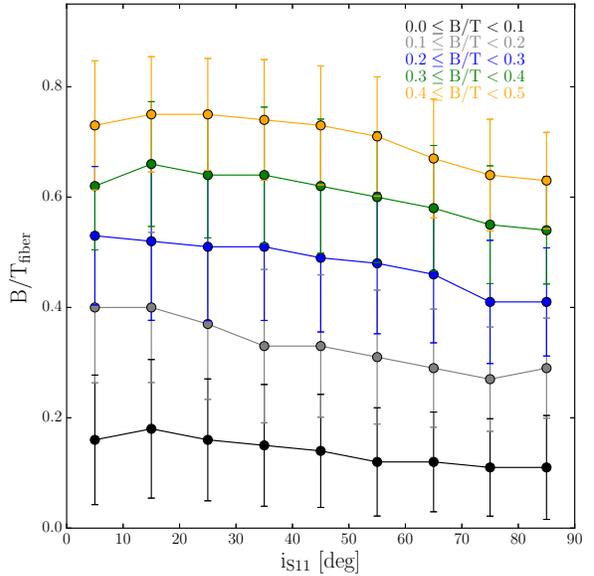}
\caption{ Fiber B/T (B/T$_{fiber}$) as a function of the disk inclination angle ($i_{S11}$) for galaxies with total stellar mass in the range 10.75 $\le$ Log$(M_{\star}/M_{\odot})<$ 11.0. Different colours mark galaxies in certain ranges of B/T, as indicated in the legend of the figure. There is a weak trend of decreasing B/T$_{fiber}$ with increasing disk inclination angle, as expected since the fraction of disk light entering the fiber is an increasing function of the disk inclination. }
\label{f_ref}
\end{figure}

\section{Discussion}

\subsection{Interpretation of the inclination effects}

Figure \ref{f6} illustrates the two major inclination effects, which grow more prominent with increasing stellar mass: with increasing disk inclination- (1) the SF peak shifts to lower SFRs, and (2) the strength of the  SF peak increases relative to the peak of quenched galaxies. 
We believe that they reflect two inclination-dependant effects that have a role in determining the total SFR of a galaxy and are dependent on the stellar mass of the galaxy. 
 The first effect is independent of morphology and is related to dust obscuration. For more inclined galaxies as photons have to travel longer distances within the disk itself before reaching the observer, an increasing fraction of the H$\alpha$ and H$\beta$ fluxes gets absorbed within the plane of the galaxy along the line of sight. As discussed in the previous Section, from Fig. \ref{f7} we can distinguish two different ranges of stellar mass where the trends in the position of the $fiber$ SFR peak with increasing disk inclinations are opposite to each other. For low mass galaxies, M$_{\star}<10^{10.25}M_{\odot}$, the peak of the SF gaussian gradually shifts to larger values when going from face-on to edge-on disks. This implies that the shift at lower values of the $total$ SF peak seen for low mass galaxies in Fig. \ref{f6} is actually caused by the aperture correction. At larger stellar masses, the aperture correction acts together with the dust content of the galaxy resulting in a shift of $\sim0.15-0.20$ dex towards lower values of the $total$ SF peak for a galaxy going from a nearly face-on disk to a nearly edge on disk. 

The second effect is dependant on galaxy morphology and it is more evident for more massive galaxies, since MS massive galaxies are characterised by larger B/T values than MS counterparts at lower stellar mass.  In particular, as we discussed earlier, we believe that this effect is due to a trend, with disk inclination, of the bulge and disk fractions that are sampled by the fibers. To show this effect, we make use of the fiber B/T ratio, B/T$_{fiber}$,  that is given in the S11 catalogue. As for the total B/T, we use here the B/T$_{fiber}$ computed from the $r-$band. In Fig. \ref{f_ref}, B/T$_{fiber}$ is shown as a function of the disk inclination angle, $i_{S11}$, for galaxies in different bins of total B/T (in different colours) and stellar mass in the range 10$^{10.75}$-10$^{11}M_{\odot}$. As expected, there is a weak trend of decreasing B/T$_{fiber}$ as the disk inclination increases, due to a larger fraction of the disk entering the fiber. This effect is small as the total B/T is fixed and the decrease in B/T$_{fiber}$ with increasing $i_{S11}$ is steeper for galaxies with larger B/T. The slope also varies with stellar mass, and it is steeper for more massive galaxies than for lower mass ones. This is because at fixed B/T the bulge component in more massive galaxies is characterised by a larger radius than the bulge of a lower mass galaxy, hence the variations with disk inclination within the fiber are stronger. In massive, face-on star-forming galaxies the $3''$ fibers sample  --in most cases--  just the bulge of the galaxies. Since most bulges are quenched most fibers sample just the quenched part of the galaxies. Apparently, in such cases the aperture correction is not sufficient to properly include the full SFR of the disk. However, as the inclination increases, an increasing fraction of the SF disk enters the fiber, the galaxy is recognised as SF and the aperture correction goes some way towards including the SFR of the disk. Still, quite possibly not all of it, as we know that the aperture correction and dust content are responsible, in massive galaxies, of a shift of $\sim0.15-0.20$ dex towards lower values of the $total$ SF peak. The shift of $\sim 0.4$ dex in the position of the SF peak in the most massive edge-on  galaxies indicates that the actual SFR is on average underestimated by a factor $\sim 2.5$ for these galaxies, a quite significant effect. 

Overall, this extreme inclination effect is relevant for only a small fraction of galaxies, say $\sim 25-30\%$ of the whole sample, and especially confined to the most massive galaxies. Therefore, by and large the  bulk of the population is reasonably well accounted for. But minorities count for some interesting issue, as discussed next.

\subsection{Effects on the Main sequence}
\begin{figure*}
 \centering
 \includegraphics[width=0.7\textwidth, keepaspectratio, trim=0cm 0cm 0cm 0cm]{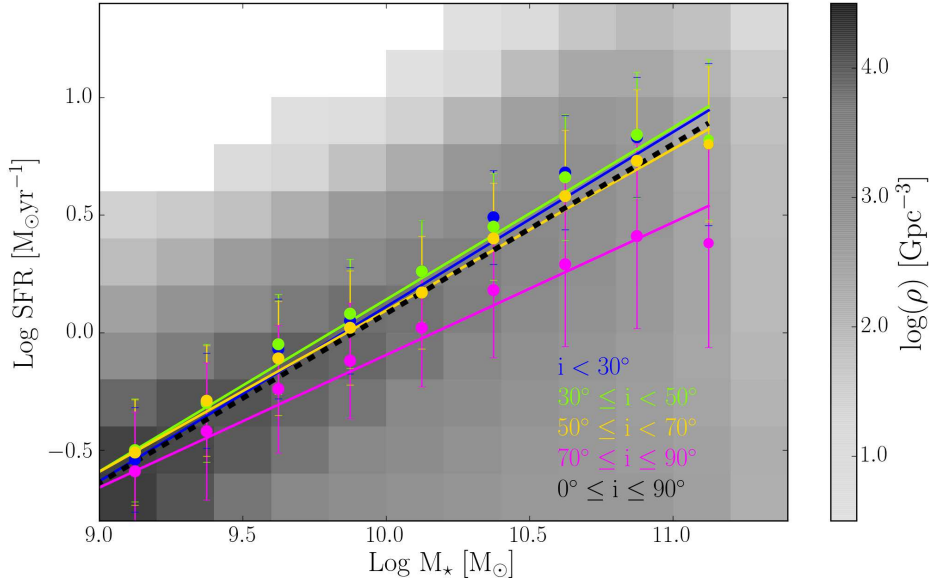}
\caption{ The SFR-$M_\star$ plane and the MS relation for galaxies that have different disk inclination angles. Each bin has been color-coded as a function of the number of galaxies divided by the average $V_{max}$ value in the bin. The dotted black line is the MS for all galaxies in our final catalogue, i.e. for galaxies with disk inclination angles $0^{\circ}\le i\le 90^{\circ}$. The MS relations are obtained from a linear regression of the mean values of the gaussian fit to the high side of the SFR distributions, as shown in Figure \ref{f5} (red gaussians). The same applies to the dispersion values, shown as vertical bars. The MS fit for the total sample, irrespective of inclination, is shown in black.}
\label{f8}
\end{figure*}

Figure \ref{f8} shows the SFR-$M_{\star}$ plane where each bin in Log(SFR) and Log($M_{\star}$) has been colour coded as a function of the space density in that bin, computed as the ratio of the number of galaxies and the average $V_{max}$, and it is indicated by the greyscale.  The best fit MS relations, i.e., Log(SFR)\ =\ $m\, {\rm Log}(M_{\star})+c$, computed in different bins of disk inclination angle, are shown in Figure \ref{f8} with different colours. In each stellar mass bin, the MS value and dispersion are the values written in red in each panel of Figure \ref{f6}, and are the peak and dispersion of the gaussian curve obtained from the fit to the right part of the SF distribution of SFR (gaussians are shown in red in Figure \ref{f6}). The slope and intercept of the MS relation in each disk inclination angle interval are obtained from a linear regression using the MS values in each stellar mass bin, covering the range ($10^{8.5}-10^{11.25}\,\msun$). The resulting  SFR-$M_{\star}$ relation for the whole sample of SF galaxies (black dotted line in Figure  \ref{f8}) has a slope of $m=0.72\pm0.02$ and an intercept of -7.12. If we limit the stellar mass range to $10^{8.5}-10^{11.0}\,\msun$, hence avoiding the most massive bin, for the whole galaxy sample we obtain $m=0.76\pm0.02$, that is the slope found by \cite{Renzini15}. For the subsample of galaxies characterised by nearly face-on disks ($i<30^{\circ}$; in blue in Figure \ref{f8}) we obtain a slope $m=0.74\pm0.02$,  with $c=-7.31$. Galaxies with disk inclination angle between $30^\circ \le i <50^\circ$ (in green) have a slope of $m=0.73\pm0.02$ and $c=-7.17$, while for disk inclination angle between and $50^\circ \le i <70^\circ$ (in yellow) the slope is smaller and the intercept is larger: $m=0.68\pm0.03$, $c=-6.76$. Finally, for nearly edge-on disks, ($i\ge70^{\circ}$; in magenta in Figure \ref{f5}) the slope is flatter, $m=0.56\pm0.03$, and the intercept larger, $c=-5.73$. 
Hence, the MS becomes progressively flatter and the intercept increases when considering galaxies with increasingly inclined disks. In addition to the change of slope and intercept of the MS for different disk inclination angles, another effect that seems clear from Figure \ref{f8} is that the lower envelope of the MS relation and the green valley region seem to be contaminated by galaxies with highly inclined disks. We analyze this effect in the next Section.

\subsection{Contaminating the Green valley}
\label{green}

The shift to lower SFRs of the SF peak of massive galaxies with increasing inclination has an adverse effect on efforts of identifying really {\it quenching} galaxies, i.e., galaxies that have left the main sequence,  their SFR is dropping, and are destined to join the {\it graveyard} of quenched galaxies. Even if the total number of high inclination galaxies is relatively small, within the green valley they may even outnumber the truly quenching galaxies.  

 \begin{figure*}
 \centering
 \includegraphics[width=0.75\textwidth, keepaspectratio, trim=0 0cm 0 1cm]{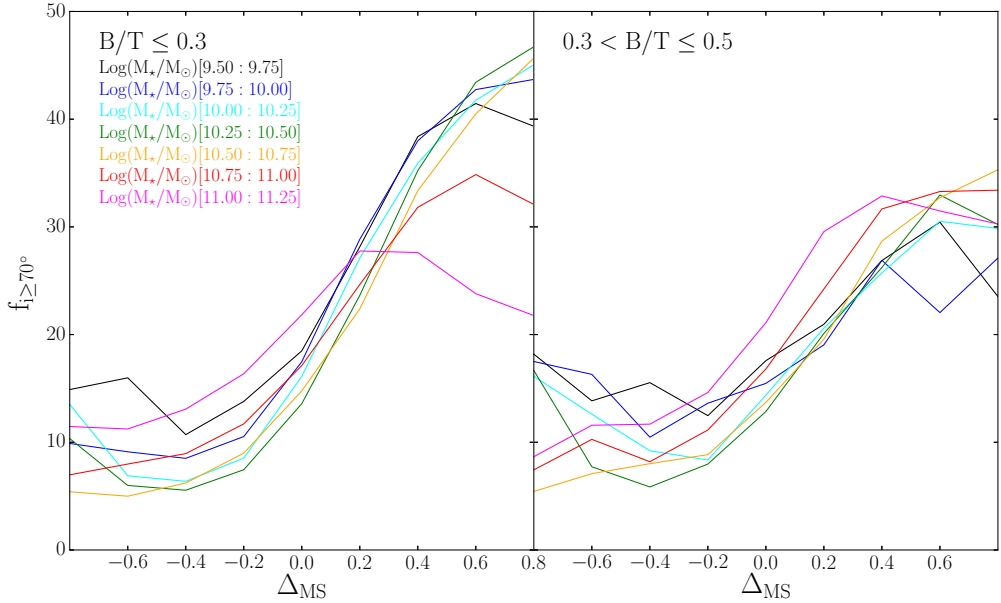}
\caption{Fraction of galaxies with disk inclination angle $i_{S11}\ge70^{\circ}$ as a function of the distance from the MS of star forming galaxies. In the $left \ panel$ the subsample of galaxies with B/T$\le$0.3 is shown, while the $right\ panel$ shows galaxies with 0.3$<$B/T$\le$0.5. The distance from the MS is computed in bins of 0.2 dex, and it is negative for galaxies above the MS, positive for galaxies below the MS. Different colours mark different stellar mass bins. There is a strong trend of increasing fraction of nearly edge-on galaxies when moving at fixed stellar mass from 0.8 dex above the MS, towards lower SFR, up to 0.8 dex below the MS. This trend is stronger for the lowest B/T galaxies. }
\label{f9}
 \end{figure*}

To quantify the contamination of high-inclined disks in the green valley, we  plot, for each stellar mass bin, the fraction of galaxies with disk inclination $i\ge70^{\circ}$, $f_{i>70^\circ}$, as a function of the Log(SFR) distance from the MS, $\Delta_{\rm MS}$. Here, the MS values are the ones obtained for the total population of galaxies, regardless of the disk inclination. $\Delta_{\rm MS}$ has been computed in bins that are 0.2 dex wide, and it is negative for galaxies above the MS, positive for galaxies below the MS. Results are shown in Figure \ref{f9}: the left panel shows galaxies with B/T$\le$0.3, while in the right panel galaxies with 0.3$<$B/T$\le$0.5 are shown, to underline trends with morphology. Each stellar mass bin is indicated by a different colour. 

There is a strong increase of the fraction of nearly edge-on galaxies, $f_{i>70^\circ}$, when moving from the upper envelope of the MS towards its lower envelop. For disk galaxies with M$_{\star}<10^{10.75}$M$_{\odot}$, $f_{i>70^\circ}$ computed 0.8 dex below the MS is 4 times larger than $f_{i>70^\circ}$ at 0.6 dex above the MS. For more massive disk galaxies, the peak value of $f_{i>70^\circ}$ is smaller, $\sim35\%$ for M$_{\star}<10^{10.75-11.0}$M$_{\odot}$ and $\sim27\%$ for M$_{\star}<10^{11.0-11.25}$M$_{\odot}$. In fact, the SFR of the most massive galaxies is derived mainly from the D4000 break as soon as we move below the MS. As shown in Figure 4, for M$_{\star}>10^{10.8}$M$_{\odot}$, the fraction of galaxies that have SFR from $H\alpha$ is $\sim$70$\%$ on the MS, and quickly drops to $\sim$20$\%$ at $\sim$0.8 dex below the MS. Since the SFR from D4000 is less affected by reddening as it is derived from a break of the continuum, the increase of $f_{i>70^\circ}$ from the upper to the lower envelop of the MS is milder for the most  massive galaxies.  
The increase in $f_{i>70^\circ}$ is less steep for galaxies with 0.3$<$B/T$\le$0.5 (right panel in Figure \ref{f9}), where $f_{i>70^\circ}$ computed at 0.8 dex below the MS is a factor from 2 to 3 larger than above MS, and it shows less variation as a function of stellar mass. We believe that the different increase in $f_{i>70^\circ}$ for galaxies with different morphology could be caused by two effects: 1) the disk inclination angle in nearly edge-on disks might be underestimated by S11 for galaxies with $0.4<$B/T$\le$0.5, that mainly populate the lower envelop of the MS, and 2) the increase in the fraction of SFR from D4000 when moving from the MS towards its lower envelop is stronger for intermediate B/T galaxies than for disk-dominated galaxies.   

It is indeed clear that, at any stellar mass, the fraction of galaxies with highly inclined disks increases when moving from the upper MS envelope toward the lower MS envelope and the green valley region. This rises two important points: 1) there is a fraction highly inclined disks {\it intruders} in the green valley that should be removed if the goal is to identify galaxies in which the quenching is ongoing, and 2) the scatter of the MS seems to be dependent on the disk inclination angle, and hence the relation could be tighter when correcting for the inclination effect. 

\subsection{Conclusions}

We have used the spectroscopic catalogue of SDSS DR7 and the bulge-disk decomposition of Simard et al. (2011) to study the effects of disk inclination on the MS of star-forming galaxies and the contamination of highly inclined disks in the green valley region. The main findings of this work can be summarized as follows:

\begin{itemize}

\item The peak of the total SFR distribution of star-forming galaxies shifts to lower values with increasing disk inclination angles. Such a shift amounts to $\sim0.2$ dex for Log$(M_{\star}/M_{\odot})<10.25$, and increases to $\sim0.4$ dex for more massive galaxies. The dispersion of the best-fit gaussian to the SF peak slightly increases with increasing disk inclination angles, but this is a small effect (at most $\sim 0.1$ dex in $\sigma$).

\item When considering the fiber SFR, in less massive galaxies we observe an opposite trend than the one that characterises the total SFR with increasing disk inclination angle. This implies that, for low stellar masses, the shift of $0.2$ dex towards lower SFR in the total SF peak is caused by the aperture correction applied in Brinchmann et al. (2004). For more massive galaxies, we observe an average shift of the SF peak of 0.2 dex towards lower SFR values when considering nearly edge-on disks as compared to galaxies with 30$^{\circ}<i<$70$^{\circ}$. For galaxies with nearly face-on disks, the peak of the SFR distribution of star forming galaxies is found at lower values than for galaxies with 30$^{\circ}<i<$70$^{\circ}$. This shift to lower fiber SFR values in massive galaxies with nearly face-on disks is due to the presence of a massive non-star-forming bulge in the the SDSS fiber. 

\item
The relative strength of the SF and passive peaks of the SFR distributions is a function of inclination, particularly at high stellar masses, with the SF peak growing more prominent with increasing inclination. This effect reflects a change in the morphology of MS galaxies, that at high stellar masses are characterised by an increase of the bulge component. For a galaxy at a given stellar mass and B/T, the relative fractions of disk and bulge within the fiber are a function on the disk inclination, with the bulge fraction that decreases for increasing disk inclination angle. 

\item The  shift in the SFR peak with increasing inclination is the result of the combination of the different effects explained above. At low stellar masses, MS galaxies are mostly pure disks, and the shift is attributed to the aperture correction that is applied to compute the total SFR from the fiber SFR. 
At larger stellar masses the dust content of galaxies increases and the final shift of the SF peak in nearly edge on disks is the result of: 1) extinction correction to the SFR being systematically underestimated in high inclination galaxies, 2) aperture correction, and 3) increase of the B/T ratio in MS galaxies.  

\item For the subsample of nearly edge-on galaxies, the MS relation is significantly flatter than for the total sample, with a slope of $m=0.56\pm0.02$, with respect to the $m=0.74\pm0.02$ for the whole sample. Also the intercept significantly differs, going from -7.31 (whole sample) to -5.37 (nearly edge-on). 

\item The fraction of galaxies with disk inclination angle $i\ge 70^{\circ}$ is a strong function of the position in the SFR-$M_{\star}$ plane with respect to the MS. In fact, the fraction of disk dominated galaxies with $i\ge 70^{\circ}$ increases by a factor of $\sim 4$ at all stellar masses when moving from 0.8 dex above the MS to 0.8 below it. When considering more bulge dominated galaxies (0.3$<$B/T$\le$0.5), the effect is still important, but the fraction of galaxies with highly inclined disks changes by a factor of $\sim 3$ from the upper to the lower envelope of the MS.

\end{itemize}

Galaxies characterised by highly inclined disks are likely to be strongly obscured,  hence their SFR estimates are more uncertain. \cite{Masters10} and \cite{Sodre13} conclude that significant amounts of dust are present in inclined spirals, and that edge-on disks are the reddest objects in the local Universe. Knowing that, many recent works which focus on the study of quenching mechanism of galaxies based on a colour selection of the galaxy samples tend to remove inclined disks to avoid their obscuration problem (see \cite{Tojeiro13}, \cite{Willett15}). \cite{Schaw14} show that the effect  of inclination is small, and that the green valley is not appreciably depopulated after correcting for dust extinction. Still, Schawinski et al. (2014) did not consider green valley galaxies with intermediate morphology, which are the relative majority in the morphological classification of Galaxy Zoo. In principle, the green valley region can provide information on the nature and duration of the process that causes the transition of a galaxy from the blue cloud to the red sequence or, in other terms, from the MS to the passive region of the SFR-$M_{\star}$ plane. For this reason, a careful estimate of the galaxy population of the green valley cannot proceed without taking the disk inclination into account, as we have shown that a large fraction of highly inclined disks  tend to fill the green valley and lie in the lower envelope of the MS relation.

\section*{Acknowledgements}
This research was supported by the DFG cluster of excellence 'Origin and Structure of the Universe'. We like to thank Stefano Berta for input on how best calculate the $V/V_{\rm max}$ corrections.
LM and AR are grateful to the Munich Institute
for Astrophysics and Particle Physics for its hospitality, when the
present project was conceived and started. AR also acknowledges hospitality and support from  the National Astronomical Observatory of Japan, Mitaka while the final version of this paper was set up. Funding support from a INAF-PRIN-2014 grant is also acknowledged. We would also like to thank Bhaskar Agarwal for useful comments on the draft.

\bsp
\label{lastpage}

\end{document}